\documentclass[letterpaper,twocolumn,10pt]{article}
\usepackage{usenix2019,epsfig,endnotes}

\usepackage{pgfplots}
\usepackage{pgfplotstable}
\pgfplotsset{compat=1.14}

\usetikzlibrary{positioning}

\usetikzlibrary{patterns}

\usepackage[hyphens]{url}
\usepackage{hyperref}

\usepackage{enumitem}

\usepackage{multirow}
\usepackage[flushleft]{threeparttable} 
\usepackage{amssymb} 
\usepackage{arydshln}

\usepackage{msc}
\drawframe{no}

\usepackage{listings}
\definecolor{nonecol}{RGB}{247,247,247}
\definecolor{unknowncol}{RGB}{99,99,99}
\definecolor{customcol}{RGB}{37,37,37}
\definecolor{lowcol}{RGB}{222,235,247}
\definecolor{medcol}{RGB}{158,202,225}
\definecolor{highcol}{RGB}{66,146,198}

\begin{document}

\date{}

\title{\Large \bf A Study of the Feasibility of Co-located App Attacks against BLE and a Large-Scale Analysis of the Current Application-Layer Security Landscape}

\author{
{\rm Pallavi Sivakumaran }\\
Information Security Group\\
Royal Holloway University of London\\
Email: pallavi.sivakumaran.2012@rhul.ac.uk
 \and
 {\rm Jorge Blasco}\\
Information Security Group\\
Royal Holloway University of London\\
Email: jorge.blascoalis@rhul.ac.uk
}

\maketitle


\subsection*{Abstract}
Bluetooth Low Energy (BLE) is a fast-growing wireless technology with a large number of potential use cases, particularly in the IoT domain. 
Increasingly, these use cases require the storage of sensitive user data or critical device controls on the BLE device, as well as the access of this data by an augmentative mobile application.
Uncontrolled access to such data could violate user privacy, cause a device to malfunction, or even endanger lives. 
The BLE standard provides security mechanisms such as pairing and bonding to protect sensitive data such that only authenticated devices can access it.
In this paper we show how unauthorized co-located Android applications can access pairing-protected BLE data, without the user's knowledge. 
We discuss mitigation strategies in terms of the various stakeholders involved in this ecosystem, and argue that at present, the only possible option for securing BLE data is for BLE developers to implement remedial measures in the form of application-layer security between the BLE device and the Android application. 
We introduce BLECryptracer, a tool for identifying the presence of such application-layer security, and present the results of a large-scale static analysis over 18,900+ BLE-enabled Android applications. 
Our findings indicate that over 45\% of these applications do not implement measures to protect BLE data, and that cryptography is sometimes applied incorrectly in those that do. 
This implies that a potentially large number of corresponding BLE peripheral devices are vulnerable to unauthorized data access. 

\section{Introduction}
Bluetooth is a well-known technology standard for wireless data transfer, currently deployed in billions of devices worldwide~\cite{Ryan:2013:Bluetooth}. 
A more recent addition to the Bluetooth standard is Bluetooth Low Energy (BLE), which differs from Classic Bluetooth in that it incorporates a simplified version of the Bluetooth stack and targets low-energy, low-cost devices.

Its focus on resource-constrained devices has made BLE highly suited for IoT applications~\cite{Elkhodr:2016:WirelessIoT}, including personal health/fitness monitoring~\cite{Gomez:2012:Overview}, asset tracking~\cite{Bisio:2016:AssetTracking}, vehicular management~\cite{Bronzi:2016:BLE}, and home automation~\cite{Karani:2016:Implementation}. 
Most of these use cases augment the functionality of the BLE device with a mobile application. 
This application may need to read or write sensitive or critical data on the BLE device (for example, glucose measurement values stored by a continuous glucose meter, or a field that controls a door's locking mechanism in a smart home security system). 
To ensure privacy and security/safety, measures should be taken to protect such data from being accessed by unauthorized entities.

The Bluetooth specification provides means for restricting access to BLE data via \textit{pairing} and \textit{bonding}, which are mechanisms for establishing an authenticated transport between two communicating devices. 
However, when multiple applications reside on a single host, as is the case with mobile devices, there is potential for a malicious application to abuse a trusted relationship between the host and the device that was initiated by an authorized application~\cite{Naveed:2014:MisBinding}.

In this work, we show how a malicious application could take advantage of the BLE communication model on Android to read and write pairing-protected data on a BLE device without the user's knowledge. 
We also show that these unauthorized applications may be able to do so while requesting minimal permissions, thereby making them appear less invasive than even an authorized application.

We discuss various strategies, in terms of the different stakeholders involved, that can be used to secure BLE data against such unauthorized access. 
We argue that in the current landscape, it is up to the BLE device/application developers to implement application-layer security to protect the data on their devices. 
We perform a large-scale static analysis of 18,929 BLE-enabled Android applications (filtered down from an original dataset of over 4.6 million applications) to determine how many of them currently employ such protection mechanisms. 
While the results vary for BLE reads vs. writes, overall they show that more than 45\% of the tested applications do not provide cryptography-based application-layer security for BLE data. 
This number rises to about 70\% for those applications that are categorized under ``Medical". 
This information, when combined with the download counts for each application, allows us to estimate a lower bound for the number of BLE devices that may be vulnerable to unauthorized data access. 

The rest of this paper is structured as follows: Section~\ref{section:background} provides an overview of key BLE concepts, particularly with regard to data access mechanisms and restrictions. 
We demonstrate unauthorized BLE data access in Section~\ref{section:androidchannels}. 
This section also discusses stakeholders and possible mitigation strategies. 
Section~\ref{section:appanalysis} details our marketplace application analysis and examines the results. 
Related work is described in Section~\ref{section:relatedwork}, and  Section~\ref{section:conclusion} provides our concluding remarks.

\section{Background} \label{section:background}
Two devices that communicate using BLE will operate in an asymmetric configuration, with the more powerful device, referred to as the \textit{central}, taking on most of the resource-intensive work. 
The resource-constrained device is termed the \textit{peripheral} and performs tasks that are designed to consume fewer resources. 

\subsubsection*{Data Access on BLE Devices} \label{subsection:bledataaccess}
BLE, unlike Classic Bluetooth, can only handle discrete data known as \textit{attributes}. 
Attributes are stored and accessed according to rules specified by the \textit{Attribute Protocol} (ATT) and the \textit{Generic Attribute Profile} (GATT), both of which are defined in the Bluetooth standard. 
There are different types of attributes, of which \textit{characteristics} are the most relevant for our analysis, as they hold the actual data of interest. 
Related characteristics are grouped into \textit{services}, which are exposed to connected devices~\cite{Bluetoothsig:2016:Specification}.

\begin{figure}
	\centering
	\begin{tikzpicture}
		\node[inner sep=0pt] (phone) at (0,0)
		{\includegraphics[width=.2\textwidth]{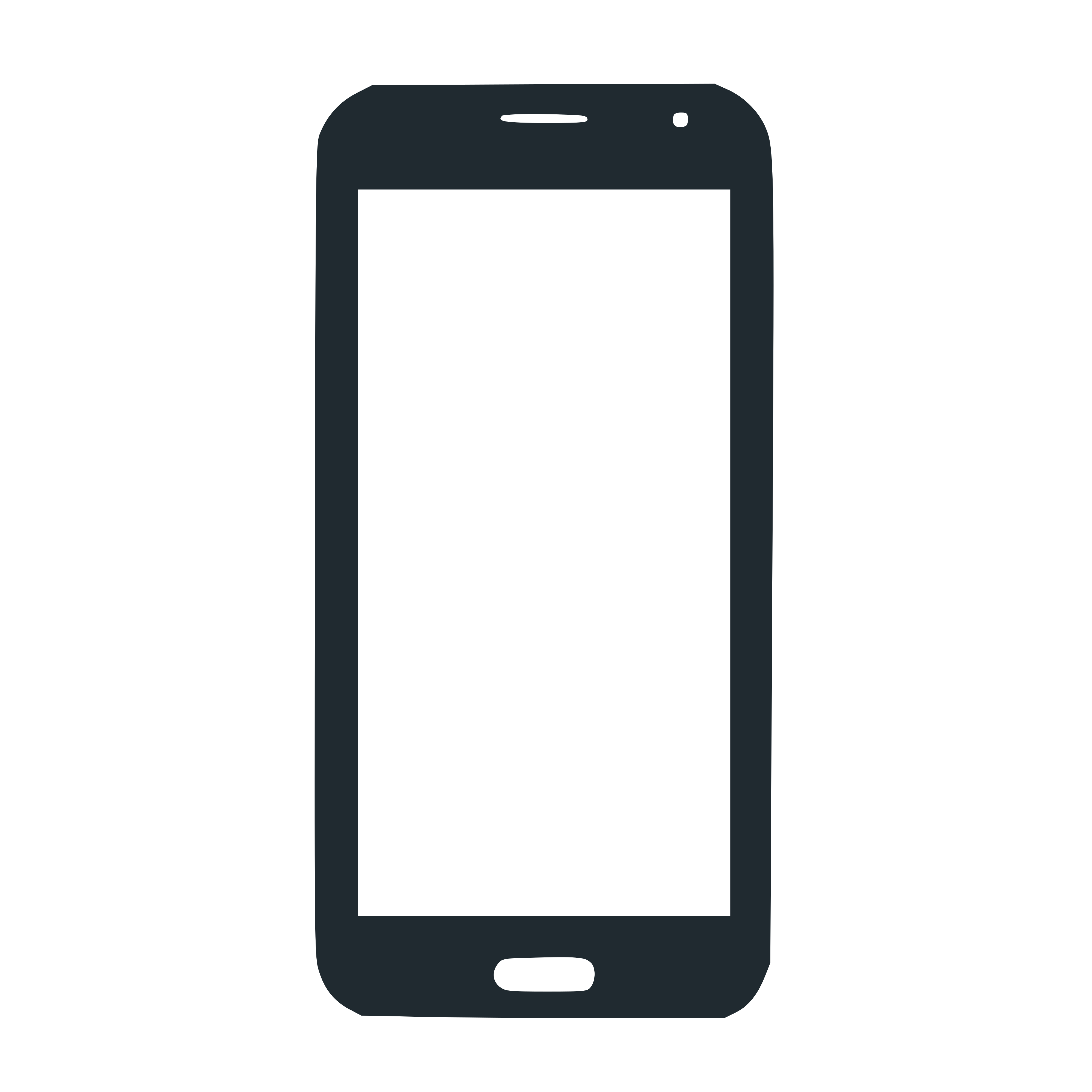}};
		\node[inner sep=0pt] (meter) at (5,0)
		{\includegraphics[width=.2\textwidth]{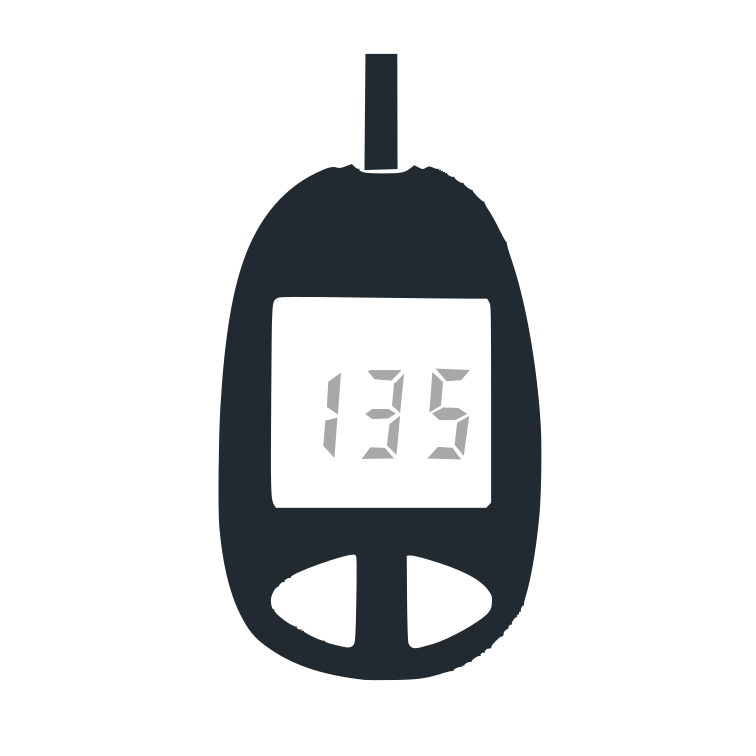}};
		\node [above=-0.1 of phone.north,anchor=south] (Central Device) {\footnotesize{CENTRAL}};
		\node [above=-0.1 of meter.north,anchor=south] (Peripheral Device) {\footnotesize{PERIPHERAL}};
		\node [below=-0.1 of phone.south,anchor=north] (gattClient) {\footnotesize{GATT Client}};
		\node [below=-0.1 of meter.south,anchor=north] (gattServer) {\footnotesize{GATT Server}};
		\draw[->] (1,0.3) -- (4,0.3) node [above,midway] {\footnotesize{read Request}};
		\draw[->] (4,-0.3) -- (1,-0.3) node [above,midway] {\footnotesize{read Response}};
	\end{tikzpicture}
	\caption{GATT communications between a mobile phone and a BLE-enabled glucometer.}
	\label{fig:blegattmsgs}
\end{figure}
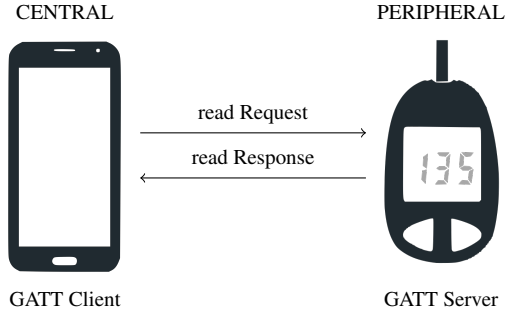

When one BLE-enabled device wants to access attributes on another BLE device, the device that initiates the exchange takes on the role of GATT \textit{client} and the other acts as the GATT \textit{server}. 
In this paper, we focus on the scenario where the BLE peripheral (e.g., a glucose meter), acts as the server, and a mobile phone acts as the client, as shown in Figure~\ref{fig:blegattmsgs}.

\subsubsection*{BLE Attribute Permissions}
Every attribute has associated with it three permissions that control how it may be accessed: (1) \textit{Access permissions} define whether an attribute can be read and/or written. (2) \textit{Authentication permissions} indicate the level of authentication and encryption that needs to be applied to the transport between the two devices before the attribute can be accessed. (3) \textit{Authorization permissions} specify whether end-user authorization is required for access.

When a GATT client sends a read or write request for an attribute to a GATT server, the server will check the request against the permissions for that attribute, to determine whether the requested access mechanism is allowed and whether the client is authenticated and/or authorized, if required. 
An attribute is only readable or writable if its access permissions specify it to be so. 
In the case of authentication permissions, if the attribute requires an authenticated or encrypted link before it can be accessed (referred to as a ``pairing-protected" attribute in this paper), and if such a link is not present when the access request is made, then the server responds with an \texttt{Insufficient Authentication/Encryption} message. 
At this point, the client can initiate the pairing process to authenticate and encrypt the transport. 
If this process completes successfully, the server will fulfill subsequent requests made by the client. 
This procedure for handling authentication requirements is well-defined in the Bluetooth specification. 
Authorization requirements, on the other hand, are implementation-specific and largely left up to developers.

Once two devices complete the pairing process, they typically go through an additional \textit{bonding} process, during which long-term keys are established. 
This prevents the need for going through the pairing process again if they disconnect and subsequently reconnect, provided they retain the long-term keys. 
Upon re-connection, the link encryption process will be initiated using the stored keys. 
Keys normally remain on the devices unless the devices are reset or manually unpaired by the user.

\section{BLE Co-Located Application Attacks} \label{section:androidchannels}
In this section, we show how \textit{any} application on an Android device can access pairing-protected attributes from a BLE peripheral, even when the pairing process was initiated by a different application. 
We then explore various mitigation strategies that are available to different stakeholders in the BLE ecosystem. 

These attacks were also explored by Naveed et al. in 2014, for Classic Bluetooth~\cite{Naveed:2014:MisBinding}. 
We show that the problem remains on newer versions of Android, and also that the situation is worse for BLE, as one of our attacks enables fewer restrictions for access and requires fewer permissions of the malicious application than even of the official application.  

\subsection{Attack Mechanisms}\label{subsection:nordicexperiments}
We describe two attacks: the first shows that pairing-protected data can be accessed by unauthorized applications, while the second refines the attack and reduces the number of permissions required by the unauthorized application.
We use two Android applications to describe the attacks: One application that is expected to be able to connect to the BLE device and access its data (``OfficialApp'') and a different application that should \textit{not} be able to access pairing-protected data from the device (``AttackApp'').

We conducted our experiments on an Alcatel Pixi 4 mobile phone, running Android 6.0 (the most widely-deployed release~\cite{Android:2018:Dashboard}), and on a Google Pixel XL, running Android 8.1.0 (the latest stable release), as of 01 Aug 2018.

\begin{figure} [!t]
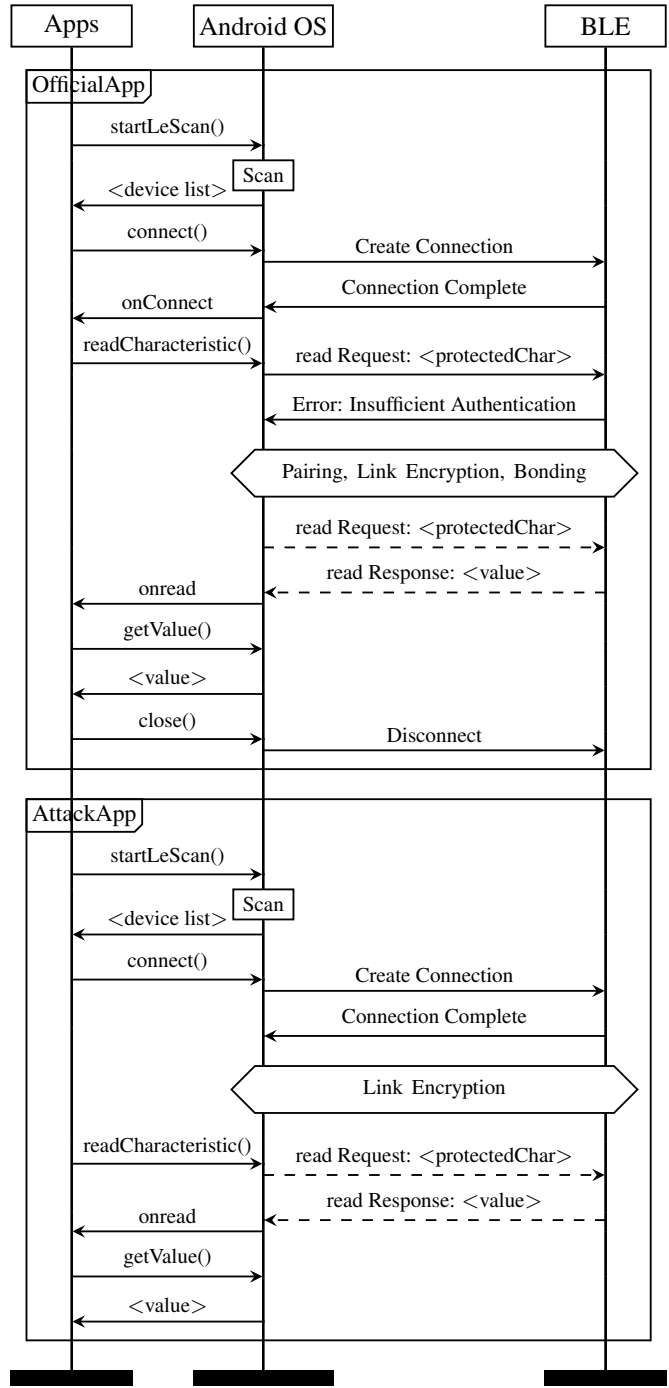

	\begin{center}
		\begin{msc}{}
			\setlength{\instdist}{0.8cm}
			\setlength{\actionwidth}{4cm}
			\setlength{\envinstdist}{0cm}
			\setlength{\msccommentdist}{0cm}
			\setlength{\conditionoverlap}{0.1cm}
			\setlength{\topheaddist}{0.2cm}
			\setlength{\bottomfootdist}{0.1cm}
			\setlength{\labeldist}{0.1cm}
			\setlength{\firstlevelheight}{0.3cm}
			\setlength{\actionwidth}{0.8cm}
			\setlength{\actionheight}{0.4cm}
			\declinst{apps}{}{Apps}
			\declinst{android}{}{Android OS}
			\setlength{\instdist}{2.8cm}
			\declinst{ble}{}{BLE}
			\inlinestart[0.6][0.6]{oa}{\small OfficialApp}{apps}{ble}
			\nextlevel[2]
			\mess{\footnotesize startLeScan()}{apps}{android}
			\nextlevel[0.4]
			\action{\footnotesize Scan}{android}
			\nextlevel[1.2]
			\mess{\footnotesize \textless{}device list\textgreater}{android}{apps}
			\nextlevel[1.2]
			\mess{\footnotesize connect()}{apps}{android}
			\nextlevel[0.3]
			\mess{\footnotesize Create Connection}{android}{ble}
			\nextlevel[1.2]
			\mess{\footnotesize Connection Complete}{ble}{android}
			\nextlevel[0.3]
			\mess{\footnotesize onConnect}{android}{apps}
			\nextlevel[1.2]
			\mess{\footnotesize readCharacteristic()}{apps}{android}
			\nextlevel[0.3]
			\mess{\footnotesize read Request: \textless{}protectedChar\textgreater}{android}{ble}
			\nextlevel[1.2]
			\mess{\footnotesize Error: Insufficient Authentication}{ble}{android}
			\nextlevel[0.8]
			\condition{\footnotesize Pairing, Link Encryption, Bonding}{android,ble}
			\nextlevel[2.6]
			\mess*{\footnotesize read Request: \textless{}protectedChar\textgreater}{android}{ble}
			\nextlevel[1.2]
			\mess*{\footnotesize read Response: \textless{}value\textgreater}{ble}{android}
			\nextlevel[0.3]
			\mess{\footnotesize onread}{android}{apps}
			\nextlevel[1.2]
			\mess{\footnotesize getValue()}{apps}{android}
			\nextlevel[1.2]
			\mess{\footnotesize \textless{}value\textgreater}{android}{apps}
			\nextlevel[1.2]
			\mess{\footnotesize close()}{apps}{android}
			\nextlevel[0.3]
			\mess{\footnotesize Disconnect}{android}{ble}
			\nextlevel[0.5]  
			\inlineend{oa}
			\nextlevel[0.8]
			  
			\inlinestart[0.6][0.6]{aa}{\small AttackApp}{apps}{ble}
			\nextlevel[2]
			\mess{\footnotesize startLeScan()}{apps}{android}
			\nextlevel[0.4]
			\action{\footnotesize Scan}{android}
			\nextlevel[1.2]
			\mess{\footnotesize \textless{}device list\textgreater}{android}{apps}
			\nextlevel[1.2]
			\mess{\footnotesize connect()}{apps}{android}
			\nextlevel[0.3]
			\mess{\footnotesize Create Connection}{android}{ble}
			\nextlevel[1.2]
			\mess{\footnotesize Connection Complete}{ble}{android}
			\nextlevel[0.8]
			\condition{\footnotesize Link Encryption}{android,ble}
			\nextlevel[2.6]
			\mess{\footnotesize readCharacteristic()}{apps}{android}
			\nextlevel[0.3]
			\mess*{\footnotesize read Request: \textless{}protectedChar\textgreater}{android}{ble}
			\nextlevel[1.2]
			\mess*{\footnotesize read Response: \textless{}value\textgreater}{ble}{android}
			\nextlevel[0.3]
			\mess{\footnotesize onread}{android}{apps}
			\nextlevel[1.2]
			\mess{\footnotesize getValue()}{apps}{android}
			\nextlevel[1.2]
			\mess{\footnotesize \textless{}value\textgreater}{android}{apps}
			\nextlevel[0.5]  
			\inlineend{aa}
		\end{msc}    
		\caption{Attack 1 - Illustrative message exchange depicting access of pairing-protected data by unauthorized application. \emph{Note:} Dashed lines indicate encrypted traffic.}
		\label{fig:mscreconnect}
	\end{center}
\end{figure}

\subsubsection*{Attack 1: System-wide Pairing Credentials} \label{subsection:attack1}
This attack demonstrates that the BLE credentials that are stored on an Android device are implicitly available to \textit{all} applications on the device, rather than just the application that originally triggered the pairing. 

When the OfficialApp connects to the BLE device and attempts to access a pairing-protected characteristic, the resulting exchange will trigger the Android OS into initiating the pairing and bonding process (as depicted in the upper block in Figure~\ref{fig:mscreconnect}). 
The resultant keys are associated with the link between the Android and BLE devices, rather than between the BLE device and the OfficialApp (which actually triggered the pairing). 
Therefore, once bonding completes, when the AttackApp scans and connects to the BLE device, the Android OS completes the connection process and automatically initiates link encryption with the keys that were generated during the previous bonding process (lower block in Figure~\ref{fig:mscreconnect}). 
This enables the AttackApp to have the same level of access to the pairing-protected data on the device as the OfficialApp, but without the need for initiating pairing.

A key point to note here is that, not only is the unauthorized AttackApp able to access potentially sensitive information from the BLE device, but also the user is likely to be unaware of the fact that this data access is taking place, as there is no indication during link re-encryption and subsequent attribute access.

\subsubsection*{Attack 2: Reuse of Connection} \label{subsection:attack2}
\begin{figure} [!t]
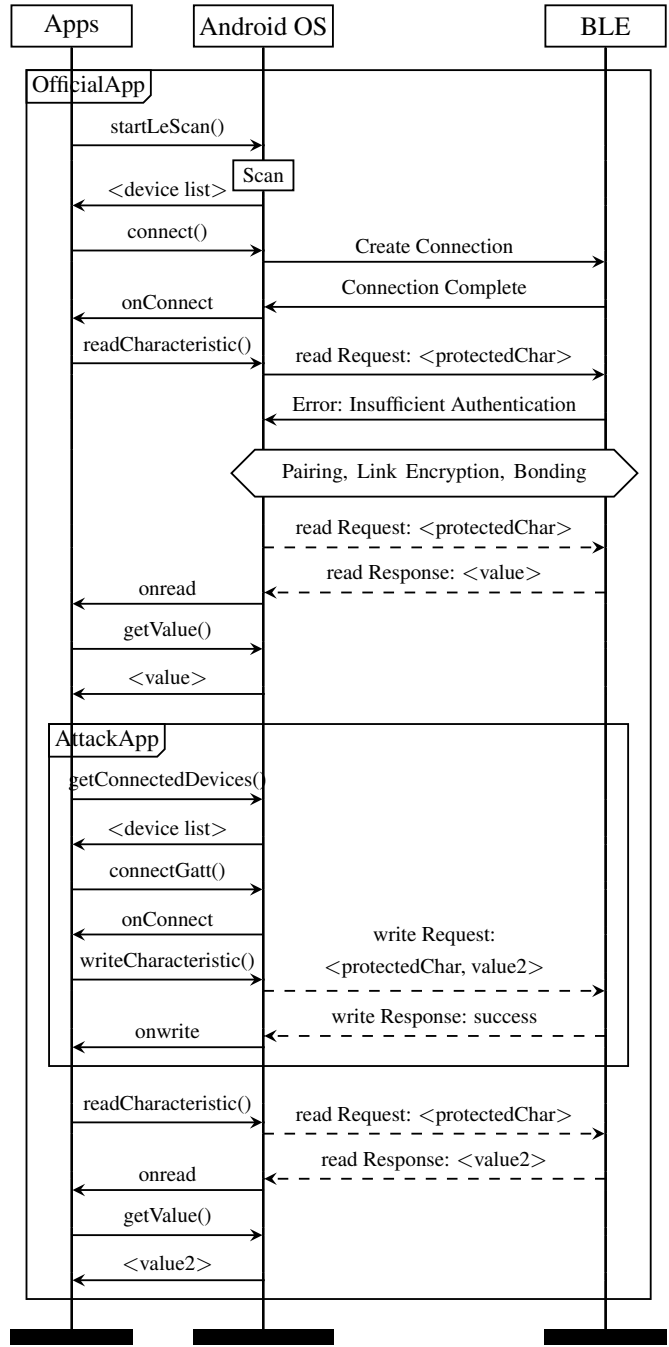

	\begin{center}
		\begin{msc}{}
			\setlength{\instdist}{0.8cm}
			\setlength{\actionwidth}{4cm}
			\setlength{\envinstdist}{0cm}
			\setlength{\msccommentdist}{0cm}
			\setlength{\conditionoverlap}{0.1cm}
			\setlength{\topheaddist}{0.2cm}
			\setlength{\bottomfootdist}{0.1cm}
			\setlength{\labeldist}{0.1cm}
			\setlength{\firstlevelheight}{0.3cm}
			\setlength{\actionwidth}{0.8cm}
			\setlength{\actionheight}{0.4cm}
			\declinst{apps}{}{Apps}
			\declinst{android}{}{Android OS}
			\setlength{\instdist}{2.8cm}
			\declinst{ble}{}{BLE}
			\inlinestart[0.6][0.6]{oa2}{\small OfficialApp}{apps}{ble}
			\nextlevel[2]
			\mess{\footnotesize startLeScan()}{apps}{android}
			\nextlevel[0.4]
			\action{\footnotesize Scan}{android}
			\nextlevel[1.2]
			\mess{\footnotesize \textless{}device list\textgreater}{android}{apps}
			\nextlevel[1.2]
			\mess{\footnotesize connect()}{apps}{android}
			\nextlevel[0.3]
			\mess{\footnotesize Create Connection}{android}{ble}
			\nextlevel[1.2]
			\mess{\footnotesize Connection Complete}{ble}{android}
			\nextlevel[0.3]
			\mess{\footnotesize onConnect}{android}{apps}
			\nextlevel[1.2]
			\mess{\footnotesize readCharacteristic()}{apps}{android}
			\nextlevel[0.3]
			\mess{\footnotesize read Request: \textless{}protectedChar\textgreater}{android}{ble}
			\nextlevel[1.2]
			\mess{\footnotesize Error: Insufficient Authentication}{ble}{android}
			\nextlevel[0.8]
			\condition{\footnotesize Pairing, Link Encryption, Bonding}{android,ble}
			\nextlevel[2.6]
			\mess*{\footnotesize read Request: \textless{}protectedChar\textgreater}{android}{ble}
			\nextlevel[1.2]
			\mess*{\footnotesize read Response: \textless{}value\textgreater}{ble}{android}
			\nextlevel[0.3]
			\mess{\footnotesize onread}{android}{apps}
			\nextlevel[1.2]
			\mess{\footnotesize getValue()}{apps}{android}
			\nextlevel[1.2]
			\mess{\footnotesize \textless{}value\textgreater}{android}{apps}
			\nextlevel[0.8]
			\inlinestart[0.3][0.3]{aa2}{\small AttackApp}{apps}{ble}
			\nextlevel[2]
			\mess{\footnotesize getConnectedDevices()}{apps}{android}
			\nextlevel[1.2]
			\mess{\footnotesize \textless{}device list\textgreater}{android}{apps}
			\nextlevel[1.2]
			\mess{\footnotesize connectGatt()}{apps}{android}
			\nextlevel[1.2]
			\mess{\footnotesize onConnect}{android}{apps}
			\nextlevel[1.2]
			\mess{\footnotesize writeCharacteristic()}{apps}{android}
			\nextlevel[0.3]
			\mess*{\begin{tabular}[t]{@{}c@{}}
				\footnotesize write Request:\\\footnotesize\textless{}protectedChar, value2\textgreater
				\end{tabular}}{android}{ble}
			\nextlevel[1.2]
			\mess*{\footnotesize write Response: success}{ble}{android}
			\nextlevel[0.3]
			\mess{\footnotesize onwrite}{android}{apps}
			\nextlevel[0.5]  
			\inlineend{aa2}
			\nextlevel[1.5] 
			\mess{\footnotesize readCharacteristic()}{apps}{android}
			\nextlevel[0.3]
			\mess*{\footnotesize read Request: \textless{}protectedChar\textgreater}{android}{ble}
			\nextlevel[1.2]
			\mess*{\footnotesize read Response: \textless{}value2\textgreater}{ble}{android}
			\nextlevel[0.3]
			\mess{\footnotesize onread}{android}{apps}
			\nextlevel[1.2]
			\mess{\footnotesize getValue()}{apps}{android}
			\nextlevel[1.2]
			\mess{\footnotesize \textless{}value2\textgreater}{android}{apps}
			\nextlevel[0.5] 
			\inlineend{oa2}
		\end{msc}
		\caption{Attack 2 - Illustrative message exchange depicting the access of pairing-protected data by reusing an existing connection. \emph{Note:} Dashed lines indicate encrypted traffic.}
		\label{fig:mscexistingconnect}
	\end{center}
\end{figure}

Our second attack exploits the fact that, on Android, a BLE peripheral can be used concurrently by multiple applications~\cite{Nordic:2016:AndroidBLE}. 
In this attack, the AttackApp does not scan for BLE devices. 
It instead searches for connected BLE devices using the \texttt{BluetoothManager.getConnectedDevices()} API call, with \texttt{BluetoothProfile.GATT} as the argument. 
If the OfficialApp happens to be in communication with the BLE device at the same time, this call will return a list with a reference to the connected BLE device. 
The AttackApp is then able to directly connect to the GATT server and read and write to the characteristics on it (including those that are pairing-protected), without the need for creating a new connection to the peripheral. 
This again is done surreptitiously, without the user being aware of the data access. 
An illustrative message flow where the AttackApp writes to a protected characteristic on the BLE device (which the OfficialApp subsequently reads) has been depicted in Figure~\ref{fig:mscexistingconnect}.

An interesting observation from this attack is a subtle but relevant impact it has on user awareness, due to the different permissions that need to be requested by the two applications. 
Since both applications access data from a GATT server, they both require \texttt{BLUETOOTH} permissions. 
In this attack scenario, because the OfficialApp scans for the BLE device before it connects to it, it also needs to request the \texttt{BLUETOOTH\_ADMIN} permission. 
Both \texttt{BLUETOOTH} and \texttt{BLUETOOTH\_ADMIN} are ``normal" permissions that are granted automatically by the Android operating system after installation, without any need for user interaction. 
However, due to restrictions imposed from Android version 6.0 onward, the OfficialApp also needs to request \texttt{LOCATION} permissions to invoke the BLE scanner without a filter (i.e., to scan for all nearby devices instead of a particular device). 
These permissions are classed as ``dangerous" and will prompt the system to display a confirmation dialog box the first time they are required. 
Because the AttackApp merely has to query the Android OS for a list of already connected devices, it does not require these additional permissions. 
This makes the AttackApp appear to be less invasive in the eyes of a user, since it does not request any permission that involves user privacy. This could play a part in determining the volume of downloads for a malicious application. 
For example, a malicious application that masquerades as a gaming application, and which does not request any dangerous permissions, may be more likely to be downloaded by end users as opposed to one that requests location permissions.

\subsection{Discussion}\label{subsection:attackdiscussion}
In this section we discuss the impact of our findings, compare them with the Classic Bluetooth case, and mention some attack limitations.

\subsubsection*{Implications of Attack}
In both of our experiments, the AttackApp was able to read and write pairing-protected data from the BLE device. 
The simplest form of attack would then be for a malicious application to perform unauthorized reads of personal user data (as an example) and relay this to a remote server.

We verified the practicability of this attack by testing a BLE-enabled fitness tracker that implemented the Bluetooth Heart Rate Service. 
The service specification states that characteristics within this service are only supposed to be protected by pairing~\cite{Bluetoothsig:2011:HRPSpecification}. 
However, we observed that the pairing employed by the device appeared to be a non-standard implementation, and also that access to the Heart Rate Measurement characteristic was ``locked" and had to be ``unlocked" by first writing to certain other characteristics on the tracker.
Despite this, we found that by deploying our second attack, our AttackApp was able to obtain Heart Rate Measurement readings without the need for performing any ``unlocking". 
This is because the AttackApp connects to the GATT server by reusing an existing connection that was initiated by the official application. The unlocking procedure would therefore already have been performed for that connection by the official application. 
This result shows that artificially restricting access to data using non-cryptographic means will not be effective. 
We notified the device developer of this issue on 01 Nov 2018, but have not yet received a response.

It should be noted that the above attack could be used by a malicious application to target other sensitive health information such as ECG, glucose or blood pressure measurements from vulnerable BLE devices, to build up a profile on a user's health. 
Further, Smart Home devices and BLE-enabled vehicles may hold information on a user's habits and lifestyle (e.g., time at home, alcohol consumption, driving speed), and could be exploited. 
It may also be possible for a malicious application to overwrite values on the BLE device, such that the written data either causes unexpected behavior on the device, or is read back by the legitimate application, thereby giving the user an incorrect view of the data on the peripheral. 
For example, it may be possible to update the peripheral's firmware via GATT writes. 
If this mechanism is not suitably protected, then a malicious application could potentially install malicious firmware onto the BLE device, as we demonstrate in Section~\ref{subsection:firmware}.

\subsubsection*{Comparison with Classic Bluetooth}
In their experiments with Classic Bluetooth, Naveed et al. found that an unauthorized Android application would not be able to obtain data from a Classic Bluetooth device if the authorized application had already established a socket connection with the device, as only one application can be in communication with the device at one time. 
Therefore, a malicious application would either require some side-channel information in order to determine the correct moment for data access, or would need to interfere with the existing connection, thereby potentially alerting the user~\cite{Naveed:2014:MisBinding}. 
This limits the attack window for the malicious application. 
Our experiments show that this is not the case with BLE communication channels. 
With BLE, there are no socket connections and if the official application has established a connection with the BLE device, then this connection can be utilized by any application that is running on the Android device. 
That is, a malicious application does not have to wait for the authorized application to disconnect before it can access data.

\subsubsection*{Attack Limitations}
The main limitation for the AttackApp in the case of the first attack is that it requires the \texttt{BLUETOOTH} and \texttt{BLUETOOTH\_ADMIN} permissions in its manifest, and also needs to explicitly request \texttt{LOCATION} permissions at first runtime in order to be able to invoke the BLE scanner. 
This enables the AttackApp to connect to the BLE device regardless of whether or not another application is also connected, but increases the risk of raising a user's suspicions. 

In the second attack scenario, the obvious limitation for the AttackApp that requests only the \texttt{BLUETOOTH} permission is that the application will only be able to access data from the BLE peripheral when the peripheral is already in a connection with (another application on) the Android device. 
That is, data access will have to be opportunistic. 
This can be achieved, for example, by periodically polling for a list of connected devices.

\subsection{Stakeholders, Mitigation Strategies and Awareness} \label{subsection:mitigation}
In this section, we discuss potential mitigating strategies that different stakeholders within the BLE ecosystem could implement in order to prevent the attacks detailed in Section~\ref{subsection:nordicexperiments}. 
We consider the Bluetooth Special Interest Group (SIG), Android (i.e., Google), and BLE device/application developers as stakeholders.

\subsubsection*{Bluetooth SIG}
The Bluetooth SIG is the group that is responsible for defining and maintaining the Bluetooth standard, which provides details on pairing, bonding and BLE attribute permissions. 
The SIG also defines various BLE services, including some that handle user health information, e.g., the Heart Rate Service and the Continuous Glucose Monitoring Service. 
The Bluetooth specifications for these services require only pairing as a protection mechanism for the characteristics that hold health-related measurements~\cite{Bluetoothsig:2011:HRPSpecification, Bluetoothsig:2015:CGMPSpecification}. 
This protection is intended to avoid man-in-the-middle attacks and eavesdropping. 
However, as shown in Section~\ref{subsection:nordicexperiments}, pairing will not prevent unauthorized Android applications from accessing the sensitive data held in these characteristics.

This issue could be avoided by modifying the Bluetooth specification and introducing specific security measures for protecting data at higher layers. 
However, this would require changes to all devices within the ecosystem, which may not be feasible due to the sheer volume of devices and applications currently available, and which could lead to fragmentation and reduced interoperability.
Despite this, we believe that developers accessing Bluetooth documentation should at least be made aware of the risks involved, and have therefore notified the SIG via their Support Request Form (17 Dec 2018). We were informed (19 Dec 2018) that the case had been assigned to the appropriate team for assessment.

\subsubsection*{Android}
Allowing all applications on an Android device to share BLE communication channels and long-term keys may well be by design, particularly since the BLE standard does not provide explicit mechanisms for selectively allowing or denying access to data based on the source application. 
This model may work in some situations, for example on a platform where all applications originate from the same trusted source. 
However, the Android ecosystem is such that, many of the applications on a device are from different and potentially untrusted sources. 
In this scenario, providing all applications with access to a common BLE transport opens up possibilities for attack, as we have demonstrated.

One option to eliminate the problem is to modify how the operating system handles BLE communication channels. 
The modification would require some form of association between BLE credentials and the application that triggers the pairing/bonding process. 
Each data access request would then be checked against the permissions associated with the requesting application. 
This approach is favoured by Naveed et al., who propose a re-architected Android framework which will create a \textit{bonding policy} when an application triggers pairing with a Bluetooth device~\cite{Naveed:2014:MisBinding}. 
This strategy has the advantage that Bluetooth devices will be protected by default from unauthorized access to their data. 
Further, assuming a suitably strong pairing mechanism is used, a minimum level of security will also be guaranteed. 
However, not only will the operating system(s) need to be modified, but also a mechanism will be required for ensuring that all users' mobile devices are updated. 
Otherwise, it is fairly likely that this measure will result in a fragmented ecosystem, with some devices running the modified operating system with protection mechanisms, and others running older versions of the OS with no protection.

Regardless of whether or not the above measure is implemented, we believe that developers should be made aware of the possibility of unauthorized applications accessing their BLE device data. 
At present, Android does not mention the issue in its Developer Guide~\cite{Android:2018:BLEoverview}. 
In fact, to the best of the authors' knowledge, there is only a single document, from a BLE chipset manufacturer, which explicitly references the fact that multiple Android applications can simultaneously use a connection to a BLE device~\cite{Nordic:2016:AndroidBLE}. Apart from this, the risks of ``system-wide pairing" have been mentioned in a specification issued by the Fast ID Online (FIDO) Alliance, without specific reference to mobile platforms~\cite{Fido:2017:BluetoothSpecification}. 

We submitted a security issue to the Android Security Team on 02 Nov 2018, focusing on the need of clear documentation so that developers are aware of the need for implementing additional protection measures if they are handling sensitive BLE data. We have received no response, except that the case has been assigned to a member of the team (15 Dec 2018).

\subsubsection*{BLE Device/Application Developers}
Despite the BLE stack containing an application layer, it could be argued that BLE is commonly viewed as a lower-layer technology for providing wireless communication capabilities, on top of which higher-layer technologies operate~\cite{Bluetoothsig:2017:MeshNetworking, Silva:2018:IoT}. 
This would result in the responsibility of securing user data being transferred from the Bluetooth SIG or Android to the BLE application/device developers. 
At any rate, this is the only mechanism available at present for protecting data against access by co-located applications.

That is, rather than relying solely on the pairing provided by the underlying operating system, developers can implement end-to-end security from their Android application to the BLE peripheral firmware. 
It may be possible to achieve such behavior via BLE authorization permissions, because even though their purpose is to specify a requirement for end user authorization, the behavior of BLE devices when encountering authorization requirements is implementation-specific. 
Most modern BLE chipsets implement authorization capabilities by intercepting read/write requests to the protected characteristics, and allowing for developer-specified validation.

One advantage of this method is that it gives the developer complete control over the strength of protection that is applied to BLE device data, as well as over the timings of security updates. 
However, leaving the implementation of security to the developer runs the risk of cryptography being applied improperly, thereby leaving the data vulnerable~\cite{Egele:2013:Empirical}. 
For existing developments, retrofitting application-layer security would mean that both an update for the Android application and a firmware update for the BLE device would be required, and there is a risk that the BLE firmware update procedure itself may not be secure~\cite{Arm:2016:Fota}. 

Due to the lack of clear guidelines regarding attribute security in both the Android Developer Guide and the Bluetooth specification, it is also possible that developers implement no security at all, due to an assumption that protection will be handled by pairing. 
In the next section, we test this assertion of a lack of developer awareness by exploring the current state of application-layer security deployments via a large-scale analysis of BLE-enabled Android applications.

\section{Marketplace Application Analysis} \label{section:appanalysis}
As evidenced by our experiments, it is fairly straightforward for any Android application to connect to a BLE device and read or write pairing-protected data. 
As discussed in Section~\ref{subsection:mitigation}, the only strategy available at present is for developers to implement application-layer security, typically in the form of cryptographic protection, between the Android application and the BLE peripheral.

\begin{table} [t]
	\centering
	\caption{APKs and Downloads per Google Play Category}
	\label{table:appcategories}
	{\renewcommand{\arraystretch}{1.2}
		\begin{threeparttable}
			\begin{tabular}{l c c} 
				\hline
				Category           & APKs [\footnotesize{packages}] & Downloads(\footnotesize{mil}) \\
				\hline
				Health \& Fitness  & 3012 [1263]                  & 344.95                        \\
				Lifestyle          & 1501 [1006]                  & 52.60                         \\
				Business           & 1489 [950]                   & 39.62                         \\
				Tools              & 1428 [891]                   & 6308.62                       \\
				Sports             & 1268 [685]                   & 17.74                         \\
				Travel \& Local    & 948 [545]                    & 31.83                         \\
				Productivity       & 526 [305]                    & 43.05                         \\
				Entertainment      & 446 [284]                    & 128.41                        \\
				Music \& Audio     & 406 [239]                    & 51.48                         \\
				Education          & 313 [225]                    & 3.35                          \\
				Shopping           & 383 [190]                    & 144.87                        \\
				Maps \& Navigation & 348 [181]                    & 33.21                         \\
				Medical            & 407 [177]                    & 5.68                          \\
				Communication      & 395 [146]                    & 755.89                        \\
				Finance            & 259 [126]                    & 96.38                         \\
				Auto \& Vehicles   & 236 [119]                    & 4.13                          \\
				Food \& Drink      & 146 [87]                     & 6.23                          \\
				Photography        & 114 [80]                     & 45.78                         \\
				Social             & 203 [77]                     & 663.43                        \\
				Other              & 746 [516]                    & 258.41                        \\
				
				\hline
			\end{tabular}
			\begin{tablenotes}
				\item [a] We make the assumption that all versions of an application fall under the same category.
				\item [b] Some APKs within the dataset are no longer available on Google Play and hence, have no corresponding category. These have not been included.
			\end{tablenotes}
		\end{threeparttable}
	}
\end{table}

In this section, we identify the proportion of applications that do \textit{not} implement such security mechanisms, to demonstrate a possible lack of awareness surrounding the issue, and to be able to estimate the number of devices that are potentially vulnerable to the types of attack shown in Section~\ref{subsection:nordicexperiments}. 

To identify the presence of application-layer security, there are two possible targets for analysis: BLE peripheral firmware or Android applications. 
At present, there is no public repository of BLE firmware, which means that the firmware would need to be obtained from the peripherals themselves. 
This would necessitate the purchase of a large number of devices and would not be financially viable. 
Further, reverse-engineering and analyzing BLE firmware is not straightforward, as the firmware image is usually a \texttt{.hex} file, which can typically only be converted to binary or assembly. 
Android APKs, on the other hand, are easier to obtain, and a number of decompilers exist that allow for conversion of APKs to a human-readable format. 

We therefore target Android applications for our analysis and perform the following: (1) obtain a substantial dataset of BLE-enabled Android APKs, (2) determine the BLE method calls and the cryptography libraries of interest, and (3) define a mechanism to determine whether BLE reads and writes make use of cryptographically processed data. 
We then apply this mechanism to our dataset and analyze the results.

\subsection{APK Dataset} \label{subsection:dataset}
We obtained our dataset from the AndroZoo project~\cite{Allix:2016:Androzoo}. 
This is an online repository that has been made available for research purposes and which contains APKs from several different application marketplaces. 
We focus on only those APKs that were retrieved from the official Google Play store, which nevertheless resulted in a sizeable dataset of over 4.6 million APKs. 
This dataset includes multiple versions for each application, as well as applications that are no longer available on the marketplace. 
We performed our analysis over the entire dataset, rather than focusing on only those APKs that are currently available on Google Play. 
This was in part because older versions of an application may still be residing on users' devices, and in part to be able to identify trends in application-layer security deployments over time.

\begin{table*} [t]
	\centering
	\caption{BLE Data Access Methods}
	\label{table:setgetmethods}
	{\renewcommand{\arraystretch}{1.2}
		\begin{threeparttable}
			\begin{tabular}{p{1.5cm} p{8cm} c c} 
				\hline
				Access                 & Method Signature\textsuperscript{a}          & \#APKs & \% of Total Methods\textsuperscript{b} \\
				\hline
				\multirow{4}{*}{Read}  & \texttt{byte[] getValue () }                                                       & 17896  & 61.58\%     \\
				                       & \texttt{Integer getIntValue (int, int)}                       & 8051   & 27.70\%     \\
				                       & \texttt{String getStringValue (int)}                                        & 2313   & 7.96\%      \\
				                       & \texttt{Float getFloatValue (int, int)}                     & 800    & 2.75\%      \\
				\hline
				\multirow{4}{*}{Write} & \texttt{boolean setValue (byte[] )  }                                         & 16198  & 70.49\%     \\
				                       & \texttt{boolean setValue (int, int, int)  }                & 5542   & 24.11\%     \\
				                       & \texttt{boolean setValue (String)}                                           & 627    & 2.73\%      \\
				                       & \texttt{boolean setValue (int, int, int, int)} & 611    & 2.66\%      \\
				\hline
			\end{tabular}
			
			\begin{tablenotes}
				\item[a]All methods are from the class \texttt{android.bluetooth.BluetoothGattCharacteristic}.
				\item[b]``\% of Total Methods'' refers to the percentage of occurrences of a particular method for a particular data access type (i.e., read or write), with respect to all methods that enable the same type of data access.
			\end{tablenotes}
		\end{threeparttable}
	}
\end{table*}

As we are only interested in those applications that perform BLE attribute access, and because such access always requires communicating with the GATT server on the BLE peripheral, the APKs were filtered by the \texttt{BLUETOOTH} permission declaration and by calls to the \texttt{connectGatt} method, which is called prior to performing any data reads or writes. 
18,929 APKs, comprising 11,067 unique packages\footnote{An Android application may have many versions, each of which will be a separate APK file (with a unique SHA256 fingerprint), but all of which will have the same package name. We use the terms ``unique applications'' or ``unique packages'' to denote the set of APKs that contain only the \textit{latest} version of each application.}, from the original set of 4,600,000+ APKs satisfy this criteria, and these formed our final dataset.

\subsubsection*{Application Categories}
Applications are categorized in Google Play according to their primary function, such as ``Productivity" or ``Entertainment", and it may be possible to gauge the sensitivity of the BLE data handled by an application based on the category it falls under. 
For example, ``Health and Fitness" applications are probably more likely to hold personal user data than ``Entertainment" applications.

The number of APKs per category has been listed in Table~\ref{table:appcategories} for our dataset. 
Approximately 23\% of the APKs (18\% of unique applications) fall under the categories of ``Health and Fitness" and ``Medical", with a cumulative download count of over 350 million. 
Note that the disproportionately high volume of downloads for the category ``Tools" is due to the Google and Google Play applications, which include BLE capabilities and are installed on most Android devices.

\subsection{Identification of BLE Methods and Crypto-Libraries} \label{subsection:cryptolibraries}
We perform our analysis against specific BLE methods and crypto-libraries. 
When considering BLE methods, we focus on those methods that involve data writes and reads. 
Such methods have been listed in Table~\ref{table:setgetmethods}, and function as the starting point for our analysis. 
For data writes, the \texttt{BluetoothGattCharacteristic} class within the \texttt{android.bluetooth} package has \texttt{setValue} methods that set the locally-stored value of a characteristic. 
This is then written out to the BLE peripheral. 
For data reads, the same class has \texttt{getValue} methods, which return data that is read from the BLE device. 
In a few APKs that we analyzed, BLE data access methods were also called from within other, vendor-specific libraries. 
However, we do not include these in our analysis as they are now obsolete.

For cryptography, Android builds on the Java Cryptography Architecture~\cite{Oracle:2018:JCA} and provides a number of APIs, contained within the \texttt{java.security} and \texttt{javax.crypto} packages, for integrating security into applications. 
While it is possible for developers to implement their own algorithms, Android recommends against this~\cite{Android:2018:SecurityTips}. 
We therefore consider only calls to these two packages as an indication of application-layer security. 

\subsection{BLECryptracer} \label{subsection:blecryptracer}
Identification of cryptographically-processed BLE data is in essence a taint-analysis problem. 
For instance, a call to an encryption method will taint the output variable that may later be written to a BLE device. 
For the purpose of this paper, when analyzing data that is read from a BLE peripheral, we consider the \texttt{getValue} variants in Table~\ref{table:setgetmethods} as sources and the cryptography API calls as sinks. 
For data that is written to the BLE device, we consider the cryptography API calls as sources and the \texttt{setValue} methods as sinks. 

There are a number of tools available for performing taint-analysis, such as Flowdroid~\cite{Arzt:2014:Flowdroid} and Amandroid~\cite{Wei:2014:Amandroid}. 
However, running a subset of our dataset of APKs through Amandroid (selected because of advantages over Flowdroid and other taint-analysis tools~\cite{Pauck:2018:Android}), we found that analysis of a single APK sometimes utilized over 10GB of RAM and took several hours to complete. 
We also found through manual analysis that many instances of cryptographically-processed data were not identified by Amandroid, especially when the BLE functions were called from third-party libraries. 
We therefore developed a custom Python analysis tool called BLECryptracer, to analyze \textit{all} calls to BLE \texttt{setValue} and \texttt{getValue} methods within an APK.

BLECryptracer is developed on top of Androguard~\cite{Desnos:2018:Androguard}, an open-source reverse-engineering tool that decompiles an Android APK and enables analysis of its components. 
Our tool traces values to/from BLE data access functions and determines whether the data has been cryptographically processed. 
To achieve this, it employs a technique for tracing register values which is sometimes referred to as ``slicing" and which has been utilized in several static code analyses~\cite{Egele:2013:Empirical, Hoffmann:2013:Slicing, Poeplau:2014:Execute}. 
It also traces fields, as well as messages passed via Intents\footnote{By matching the \texttt{Extra} identifier within the calling method.} and certain threading functions, e.g., \texttt{AsyncTask}.
It returns TRUE at the first instance of cryptography that it encounters and FALSE if it is unable to identify any application-layer security with BLE data.

\begin{table*} [t]
	\centering
	\caption{Accuracy Statistics}
	\label{table:benchmarkingconfidence}
	{\renewcommand{\arraystretch}{1.2}
		\begin{threeparttable}
			\begin{tabular}{lll|ccccccccc}
\hline
Access  & Tool  & Confidence & App Set\textsuperscript{a}  & Detected\textsuperscript{b}  &  TP  &  FP  &  TN  &  FN  &  Precision & Recall & F-measure \\
\hline
\multirow{4}{*}{Read} & Amandroid & N/A & 92 & 49 & 44 & 5 & 10 & 33 & 90\% & 57\% & 70\% \\
\cdashline{2-12}
 & \multirow{3}{*}{BLECryptracer} & High & 92 & 62 & 58 & 4 & 11 & 19 & 94\% & 75\% & 83\% \\
 &  & Medium & 30 & 11 & 7 & 4 & 7 & 12 & 64\% & 37\% & 47\% \\
 &  & Low & 19 & 12 & 8 & 4 & 3 & 4 & 67\% & 67\% & 67\% \\
\hline
\multirow{4}{*}{Write} & Amandroid & N/A & 92 & 56 & 49 & 7 & 8 & 28 & 88\% & 64\% & 74\% \\
\cdashline{2-12}
 & \multirow{3}{*}{BLECryptracer} & High & 92 & 50 & 46 & 4 & 11 & 31 & 92\% & 60\% & 72\% \\
 &  & Medium & 42 & 22 & 19 & 3 & 8 & 12 & 86\% & 61\% & 72\% \\
 &  & Low & 20 & 10 & 5 & 5 & 3 & 7 & 50\% & 42\% & 45\% \\
\hline

\end{tabular}
			\begin{tablenotes}
			    \item [a] Number of APKs tested. Note that, for confidence levels Medium and Low, we don't consider the APKs detected at higher confidence levels.
				\item [b] The number of APKs that were identified as having cryptographically protected BLE data.
			\end{tablenotes}
		\end{threeparttable}
	}
\end{table*}

Our tool analyzes BLE reads and writes separately, as the direction of tracing is different in the two cases. It performs three main types of tracing, in the following order:
\begin{enumerate}[noitemsep]
    \item Direct trace - Attempt to identify link between BLE and cryptography functions via direct register value transfers and as immediate results of method invocations.
    \item Associated entity trace - If the direct trace does not identify a link between source and sink, analyze abstract/instance methods and other registers used in previously analyzed function calls.
    \item ``Lenient" trace - If the above methods fail to return a positive result, perform a search through all previously encountered methods (which would have originated from the BLE data access method), to determine if cryptography is used \textit{anywhere} within them.
\end{enumerate}

The first trace method will produce results that are most likely to actually have cryptographically-processed BLE data, as the coarse-grained analysis performed in the subsequent methods adds increasing amounts of uncertainty. For this reason, BLECryptracer assigns ``confidence levels" of High, Medium and Low to its output, which correspond to the three trace methods above, to indicate how certain it is of the result. We evaluate these confidence levels against a modified version of the DroidBench benchmarking suite in Section~\ref{subsection:accuracy}. 
Note that BLECryptracer only looks for application-layer security in benign applications, and these confidence levels apply only when deliberate manipulations (i.e., malicious obfuscation techniques) are not employed to hide the data flow between source and sink. 

Appendix A describes the tracing mechanism in greater detail, and also outlines how BLECryptracer combats the effects of obfuscation in benign applications. 

\subsection{Evaluation}\label{subsection:testing}
We evaluated BLECryptracer, in terms of both accuracy and execution times. For comparison purposes, we have included test results from Amandroid as well.

\subsubsection*{Accuracy Measures} \label{subsection:accuracy}
At present, there is no dataset of real-world APKs with known use of cryptographically-processed BLE data, i.e., ground  truth. 
Therefore, in order to test our tool against different data transfer mechanisms, we re-factored the DroidBench benchmarking suite~\cite{Fritz:2018:DroidBench} for the BLE case. 

Each DroidBench test application was cloned twice and modified so the data flow between the sources and sinks would be from \texttt{getValue} to a cryptography method invocation, and from the cryptography method invocation to \texttt{setValue}, to emulate cryptographically-processed reads and writes, respectively. Some DroidBench test cases were excluded as they were found to be irrelevant due to differences in the objectives of DroidBench and our test set, e.g., applications that employ emulator detection or which leak contextual information in exceptions. Further, applications where BLE data is written to or read from files, or which contain data leaks in inactive code segments were not included (as our aim is to determine whether or not BLE data is cryptographically-processed). In total, we created 184 APKs: 92 for reads and 92 for writes.


We executed BLECryptracer against our benchmarking test set, analyzed the results and obtained performance metrics in terms of the three different confidence levels. 
The statistics differ based on the type of access that is analyzed (i.e., reads vs. writes) due to differences in the tracing mechanisms.
The same test set was also used against Amandroid for comparison. Table~\ref{table:benchmarkingconfidence} presents the performance metrics for both tools.

In the case of BLECryptracer results, the metrics are with respect to the total analyzed APKs at each confidence level. 
That is, because lower confidence levels analyze only those APKs that do not get detected at higher levels, accurate metrics can only be derived by considering the set of APKs that were actually analyzed at each level. 
For example, when considering the analysis of BLE reads, while the entire dataset of 92 APKs is relevant for confidence level High, only the 30 APKs that do \textit{not} result in a TRUE outcome at level High will be analyzed for confidence level Medium. 
This also means that, when obtaining performance metrics for confidence level High, all TRUE results obtained at levels Medium and Low are taken to be FALSE.

The DroidBench test set, and hence our benchmarking suite, is an imbalanced dataset, containing far more samples \textit{with} leaks (77) than \textit{without} (15). For this reason, metrics such as accuracy are not suitable for analyzing the performance of our tool when executed against this test set, as they are more susceptible to skew~\cite{Guo:2008:ClassImbalance, Jeni:2013:Imbalance}.  For our analysis, we compare the combined True Positive Rate (TPR) and False Positive Rate (FPR), and the combined precision-recall instead, in-line with taint-analysis evaluations~\cite{Qiu:2018:Analyzing}. 

 Table~\ref{table:benchmarkingconfidence} presents the precision and recall (i.e., TPR) for both BLECryptracer and Amandroid. We further derive FPRs for both tools. With BLECryptracer, when analyzing reads, False Positive Rates steadily increase as the confidence level reduces, as expected, with values of 27\% for confidence level High, 36\% for Medium and 57\% for Low. When analyzing writes, the values are 27\%, 27\% and 63\%, respectively. Regardless of the data access mechanism being tested, BLECryptracer (considering only the results at High confidence, for a fairer comparison) performs better than Amandroid in terms of FPR, with 27\% vs. 33\% for reads and 27\% vs. 47\% for writes. Precision values are also better in the case of BLECryptracer for both reads and writes. In terms of the True Positive Rate, BLECryptracer performs better than Amandroid for reads at 75\% vs. 57\%, and slightly worse for writes at 60\% vs. 64\%. These results show that, overall, BLECryptracer performs better than Amandroid for analyzing the use of cryptography with BLE data.

It should be noted that three of the four False Positives obtained by BLECryptracer at the High confidence level were due to the order in which variables are assigned values (i.e., lifecycle events), which is not tested for by BLECryptracer. Other data transfer mechanisms not tested for are \texttt{Looper} and \texttt{Messenger} functions, which generate False Negatives.
The remaining False Positive was due to the presence of method aliasing and was also identified as a False Positive by Amandroid.
In addition, the unexpectedly low TPR (i.e., recall) at level Medium for reads is due to the relatively few cases analyzed at that level when compared to High.


\begin{figure}
	\centering
	\begin{tikzpicture}
		\node[inner sep=0pt] (piechart) at (0,0)
		{\includegraphics[width=\linewidth,height=\textheight,keepaspectratio]{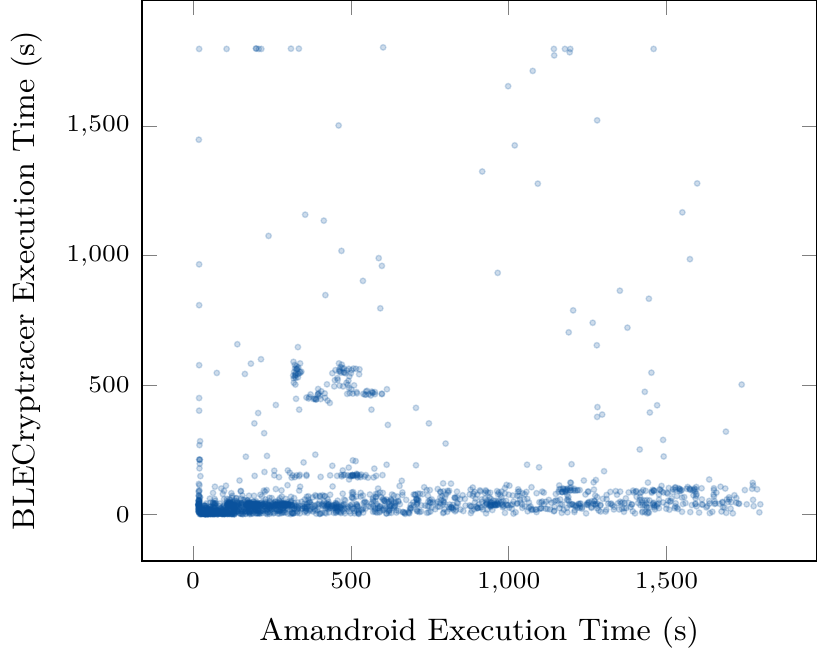}};
	\end{tikzpicture}
	\caption{Comparison of time taken to execute Amandroid vs. BLECryptracer, when analyzing BLE writes.}
	\label{fig:setvaluetimes}
\end{figure}

\subsubsection*{Execution Times}
We also compared BLECryptracer and Amandroid in terms of speed of execution. For this, we ran the two tools against a random subset of 2,000 APKs and compared time-to-completion in both cases. We imposed a maximum run-time of 30 minutes per APK for both tools, and only compared execution times for those cases where Amandroid did not time out (approximately 40\% of the tested APKs timed out when analyzed by Amandroid. In comparison, fewer than 2\% of APKs timed out when analyzed by BLECryptracer).

Figure~\ref{fig:setvaluetimes} plots the time taken to analyze BLE writes using BLECryptracer vs. Amandroid. 
The figure shows that analysis times with BLECryptracer were, for the most part, around 3-4 minutes per application. We observed no obvious correlation between the size of the application's dex file and the execution time, for either tool.
APKs that took longer to process with BLECryptracer were predominantly of confidence level ``Medium", which indicates that the longer analysis times may simply have been because of having to first go through the most stringent analysis (at the highest confidence level). 
For Amandroid, the execution times vary to a greater extent than with BLECryptracer, due to the difference in the mechanisms employed for performing the analysis.

\subsection{Results from Large-Scale APK Analysis} \label{subsection:coderesults}
We executed BLECryptracer against our dataset of 18,929 APKs. 376 APKs timed out when analyzing reads and 335 APKs timed out when analyzing writes, when a maximum runtime of 30 minutes was imposed. These APKs were re-tested with an increased runtime of 60 minutes. However, even with the longer analysis time, 114 and 161 APKs timed out for reads and writes, respectively, and had to be excluded from further analysis. In addition, 88 APKs could not be processed via Androguard's \texttt{AnalyzeAPK} method and were excluded.

Due to the differences in performance metrics obtained for the three confidence levels during testing (as mentioned in Section~\ref{subsection:testing}), we focus on only those results that either identify cryptography at confidence level High or those where no cryptography was identified at all. 

\subsubsection*{Presence of Application-Layer Security with BLE Data}
\begin{figure} [t]
	\centering
\begin{tikzpicture}
		\node[inner sep=0pt] (bargraph) at (0,0)
		{\includegraphics[width=\linewidth,height=\textheight,keepaspectratio]{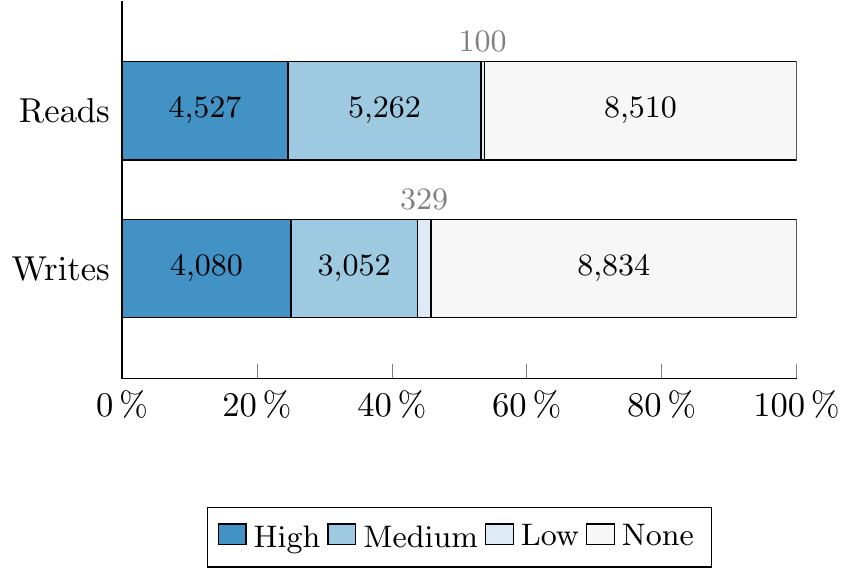}};
	\end{tikzpicture}
	\caption{Analysis results depicting the presence of cryptographically-processed data with BLE writes and reads, with breakdown according to Confidence Level.}
	\label{fig:analysisresultspie}
\end{figure}

Our results show that approximately 95\% of BLE-enabled APKs call the \texttt{javax.crypto} and \texttt{java.security} cryptography libraries \textit{somewhere} within their code. 
While this is a large proportion of APKs, the results also indicate that a much smaller percentage of APKs use cryptographically processed data \textit{with BLE reads and writes} (approximately 25\% for both, identified with High confidence). 
In fact, about 46\% of APKs that perform BLE reads and 54\% of those that perform BLE writes (corresponding to 2,379 million and 2,075 million cumulative installations, respectively) do \textit{not} implement security for the BLE data. 
Interestingly, of the 15,986 APKs that called both BLE read and write functions, about 36\% (i.e., more than 5,700 APKs), with a cumulative download count of 1,005 million, do not implement application-layer security for either type of data access. 
Figure~\ref{fig:analysisresultspie} summarizes the proportion of APKs that were identified as containing cryptographically protected BLE data at the three different confidence levels.

\subsubsection*{Third-Party Libraries vs. App-Specific Implementations}
We found that many BLE-enabled APKs actually use third-party libraries for incorporating BLE functionality.
To get an idea of exactly how many APKs relied on libraries, we analyze all methods within an APK that called BLE data access functions. 
To do this in an automated way, we compare the method class name with the application package name. If the first two components (e.g., \texttt{com.packagename}) of each match, then we take it to be a method implemented within the application. If the components do not match, we take it to be a library method. 

Of the APKs that called the \texttt{setValue} method, 63\% used BLE functionality solely through libraries, 32\% APKs used application-specific methods only, while 4\% APKs used both. Fewer than 1\% of the APKs could not be analyzed due to very short method names. Within the APKs that used both application-specific methods and libraries, around 34\% used an external library to provide Device Firmware Update (DFU) capabilities, thereby enabling the BLE peripheral to be updated via the mobile application.  Of the APKs that utilized only application-specific methods to incorporate BLE functionality, 67\% did \textit{not} implement application-layer security with the BLE data. This proportion was lower at 48\% for applications that relied on libraries.  

In the case of the APKs that called \texttt{getValue} variants, 37\% APKs used only application-specific methods, 58\% used only libraries, and 5\% used both. Similar to the \texttt{setValue} case, a higher proportion of APKs that implemented BLE functionality solely within the application did not use cryptography (60\%), when compared with those that used only libraries (39\%).

Table~\ref{table:externalblelibs} presents the ten most commonly-encountered BLE libraries, their functionality, the number of APKs that use them, and the presence of cryptographically-processed BLE data within the library itself. 
The table shows that the most prevalent third-party packages are libraries that enable communication with BLE beacons. 
In fact, a single such library (Estimote) made up more than 90\% of all instances of cryptographically-processed BLE writes and 85\% of cryptographically-processed BLE reads (identified with High confidence). 
An analysis of this library suggested that cryptography is being used to authenticate requests when modifying settings on the beacon. 

Apart from beacon libraries, we identified five libraries that function as wrappers for the Android BLE API. 
For example, Polidea wraps the API so that it adheres to the reactive programming paradigm. 
The libraries Randdusing, Megster and Evothings enable the use of BLE via JavaScript in Cordova-based applications. 
Similarly, Chromium enables websites to access BLE devices via JavaScript calls. 
None of the libraries handle cryptographically-processed BLE data. 
It is expected that developers using these libraries will implement their own application-layer security (using either JavaScript or reactive Java as appropriate).

Of the two remaining libraries, Flic, which uses cryptographically-processed data, is a library offered by a BLE device manufacturer. 
This library allows third-party developers to integrate their services into the Flic ecosystem, to allow them to automate certain tasks.

Finally, Nordicsemi is a library provided by a BLE chipset manufacturer to enable DFU over the BLE interface. 
With the newest version of the DFU mechanism, the BLE peripheral verifies that the firmware has been properly signed. 
Devices using the legacy DFU mechanism will not verify the firmware. 
However, the mobile application (and by extension, the library) does not need to handle cryptographically-processed data in either case.

\begin{table} [t]
	\centering
	\caption{Top Ten Third-Party BLE Libraries}
	\label{table:externalblelibs}
	{\renewcommand{\arraystretch}{1.2}
		\begin{threeparttable}
			\begin{tabular}{l l c c} 
				\hline
				Library    & Function       & \#APKs\footnotesize{[unique]} & Crypto \\
				\hline
				Estimote   & Beacon         & 3980[2804]                    & Yes    \\
				Nordicsemi\textsuperscript{a} & DFU            & 1238[847]                 & No     \\
				Kontakt    & Beacon         & 1108[690]                     & No     \\
				Chromium   & Web BLE        & 402[269]                      & No     \\
				Randdusing & Cordova Plugin & 268[188]                      & No     \\
				Megster    & Cordova Plugin & 317[187]                      & No     \\
				Flic       & BLE Accessory  & 173[164]                      & Yes    \\
				Polidea    & BLE Wrapper    & 138[114]                      & No     \\
				Evothings  & Cordova Plugin & 142[84]                       & No     \\
				Jaalee     & Beacon         & 102[79]                       & No     \\
				\hline
			\end{tabular}
			\begin{tablenotes}
				\item [a] Significant overlap present between Estimote and Nordic due to repackaging of the Nordic SDK into Estimote.
			\end{tablenotes}
		\end{threeparttable}
	}
\end{table}

\subsubsection*{Cryptographic Correctness}
BLECryptracer identified 3,228 unique packages with cryptographically protected BLE data (with either reads or writes), with High confidence. 
However, the presence of crypto-libraries does not in itself indicate a secure system. 
We therefore further analyzed this subset of APKs to identify whether cryptography had been used correctly in them. 
The tool CogniCrypt~\cite{Kruger:2017:Cognicrypt} was utilized for this purpose. 
Although this tool does not formally verify the cryptographic protocol between the application and the BLE peripheral, it identifies various misuses of the Java crypto/security libraries. 

Even among the 3,228 unique packages, we found that there was significant overlap between APKs in terms of the BLE libraries or functions used\footnote{There are instances where two applications may have unique package names, but which actually incorporate much the same functionality. This is often the case when the same developer produces branded variants of an application for different clients in a single industry.}. 
Removing such duplicates resulted in a set of 194 APKs.
Of these, 68 were identified by CogniCrypt as having issues. However, because CogniCrypt identifies cryptography misuse within the entire APK, the results were filtered for BLE-specific functions. 
24 APKs were found to have issues within or associated with the methods that cryptographically processed BLE data (as identified by BLECryptracer) and often, a single APK  exhibited multiple issues.
Table~\ref{table:badcrypto} shows the different types of misuse encountered and the number of unique packages that were identified as having such misuse. 
Note that because this analysis was performed over unique packages, the number of \textit{APKs} that misuse crypto-libraries will be higher. 

We manually analyzed the 24 APKs that were flagged by CogniCrypt as having BLE-relevant issues, and examined the identified instances of bad cipher modes and hardcoded keys/Initialization Vectors (IVs). With regard to insecure block cipher modes, we found that explicit use of ECB was prevalent (9 out of 10 cases), but there was also one case where \texttt{Cipher.getInstance("AES")} was used without the mode being specified, which may default to ECB depending on the cryptographic provider. When analyzing keys, we observed that several applications directly contained hardcoded keys as byte arrays or strings. Three applications retrieved keys from JSON files. In two cases, keys were generated from the \texttt{ANDROID\_ID}, which is a system setting that is readable by all applications. We also observed one instance where a key was obtained from a server via HTTP (not HTTPS). 

This analysis shows that several real-world applications contain basic mistakes in their use of crypto-libraries and handling of sensitive data, which means that the BLE data will not be secure despite the use of cryptography.

\subsubsection*{Trends over Time}
\begin{figure}
	\begin{tikzpicture}
		\begin{axis}[
				x tick label style={
					/pgf/number format/1000 sep=},
				xlabel={Year},
				xtick pos=left,
			    ymajorgrids,
			    ymax=100,
				ylabel={\% APKs with no Cryptography},
				enlargelimits=0.15,
				tick label style={font=\small},
                label style={font=\small},
                legend style={font=\small},
				legend style={at={(0.5,-0.2)},
					anchor=north,legend columns=-1},
				legend image code/.code={%
                    \draw[#1, draw=none] (0cm,-0.1cm) rectangle (0.15cm,0.25cm);
                },
				ybar,
				bar width=11pt,
			]
			\addplot[black!80,solid,fill=medcol]
			coordinates {(2013,64) (2014,51)
			(2015,42) (2016,42) (2017,35) (2018,68)};

		\end{axis}
	\end{tikzpicture}
	\caption{Application-layer security trends over time. \textit{Notes:} Graph depicts APKs that perform BLE reads or writes, and have no crypto for either. APKs with dates that are invalid~\cite{Androzoo:2016:Lists} or older than 2012 (when native BLE support was introduced for Android) have not been included.}
	\label{fig:appsecurityovertime}
\end{figure}
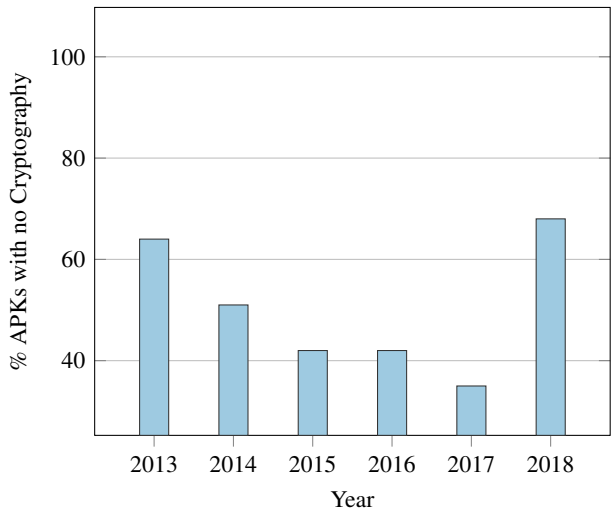

Figure~\ref{fig:appsecurityovertime} shows the trend of application-layer security over time for applications that incorporates calls to BLE reads or writes. 
The graph depicts the percentage of applications that do not have cryptographic protection for either type of access. 
The overall downward trend suggests that there has been some improvement in application-layer security between the years 2014 and 2017 (we refrain from making observations about APKs from 2013 as they were very few in number, and about APKs from 2018 as the dataset is not yet fully populated for this year). However, it should be noted that, even in 2017, which had the smallest percentage of APKs without cryptography, these APKs corresponded to 128 million downloads, which is a significant number.

\begin{table} [t]
	\centering
	\caption{Number of Packages with Cryptographic Misuse}
	\label{table:badcrypto}
	{\renewcommand{\arraystretch}{1.2}
		\begin{threeparttable}
			\begin{tabular}{l c} 
				\hline
				Misuse Type\textsuperscript{a}                      & \#Unique Packages \\
				\hline
				ECB (or other bad mode)          & 10               \\
				Non-random key                   & 6                \\
				Non-random IV                    & 10                 \\
				Bad IV used with Cipher          & 7                 \\
				Bad key used with Cipher         & 11                \\
				Incomplete operation (dead code) & 4                \\
				
				\hline
			\end{tabular}
			\begin{tablenotes}
				\item [a] Description of misuse based on~\cite{Egele:2013:Empirical,Kruger:2017:CryptoAnalysis}.
			\end{tablenotes}
		\end{threeparttable}
	}
\end{table}

\subsubsection*{Application-Layer Security by Category}
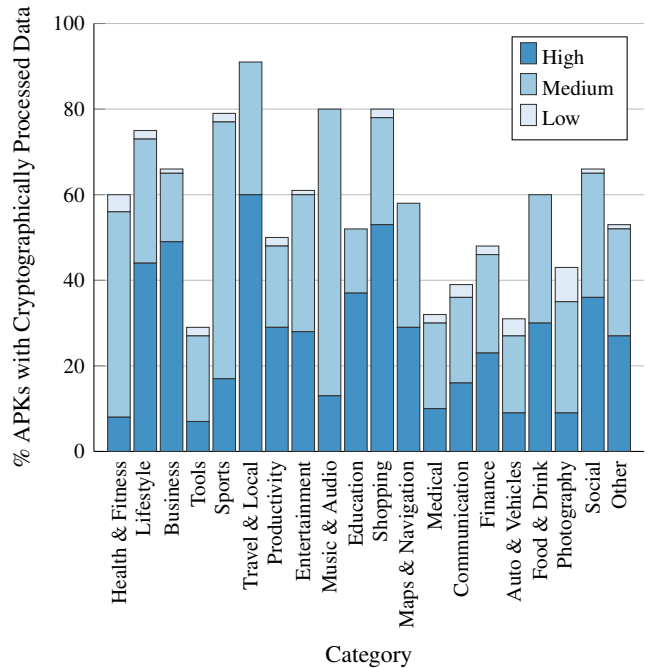
\begin{figure}[t]
	\begin{tikzpicture}
		\pgfplotstableread[col sep=comma]{
			Category,High,Medium,Low
			Health \& Fitness,8,48,4
            Lifestyle,44,29,2
            Business,49,16,1
            Tools,7,20,2
            Sports,17,60,2
            Travel \& Local,60,31,0
            Productivity,29,19,2
            Entertainment,28,32,1
            Music \& Audio,13,67,0
            Education,37,15,0
            Shopping,53,25,2
            Maps \& Navigation,29,29,0
            Medical,10,20,2
            Communication,16,20,3
            Finance,23,23,2
            Auto \& Vehicles,9,18,4
            Food \& Drink,30,30,0
            Photography,9,26,8
            Social,36,29,1
            Other,27,25,1

			}\categorytable
		
		\begin{axis}[
				x=0.35cm,
				bar width=0.30cm,
				enlarge x limits=0.05,
				ymajorgrids,
				xlabel={Category},
				ylabel={\% APKs with Cryptographically Processed Data},
				ymin=0, ymax=100,
				ybar stacked,
				symbolic x coords={Health \& Fitness,Lifestyle,Business,Tools,Sports,Travel \& Local,Productivity,Entertainment,Music \& Audio,Education,Shopping,Maps \& Navigation,Medical,Communication,Finance,Auto \& Vehicles,Food \& Drink,Photography,Social,Other
				}, 
				xtick=data, 
				x tick label style={rotate=90, /pgf/number format/1000 sep=},
				tick label style={font=\footnotesize},
				label style={font=\small},
				legend style={font=\footnotesize},
				legend pos=north east,
				legend cell align={left},
				axis y line*=none,
                axis x line*=bottom,
			]
			\addplot [black!80,fill=highcol] table[x=Category,y=High] {\categorytable};\addlegendentry{High}
			\addplot [black!80,fill=medcol] table[x=Category,y=Medium] {\categorytable};\addlegendentry{Medium}
			\addplot [black!80,fill=lowcol] table[x=Category,y=Low] {\categorytable};\addlegendentry{Low}
		\end{axis}
	\end{tikzpicture}
	\caption{Presence of application-layer security in different categories of applications, averaged over BLE reads and writes, and broken down by confidence level. Only unique packages have been taken into consideration. APKs that do not currently have a presence on Google Play have been excluded, as their category cannot be identified.}
	\label{fig:appsecuritycategories}
\end{figure}

The percentage of applications that use cryptographically processed data from each major application category has been graphed in  Figure~\ref{fig:appsecuritycategories}. 
While it would be reasonable to expect that most ``Medical" applications would implement some level of application-layer security, the results show that fewer than 30\% of applications under this category actually have such protection mechanisms. However, it is possible that the reason for this is that the devices implement the standard Bluetooth SIG adopted profiles, which do not mandate any security apart from pairing, as mentioned in Section~\ref{subsection:mitigation}. In fact, of the APKs categorized under ``Medical" and with no cryptographic protection for either reads or writes, we found that three of the top ten (in terms of installations) contained identifiers for the standard Bluetooth Glucose Service.

Perhaps surprisingly, APKs that are categorized under ``Business'', ``Shopping'' and ``Travel \& Local'' appear to be the most likely to incorporate application-layer security, with around 50\% of all such applications being identified as having cryptographically processed BLE data with High confidence. 
However, in over 85\% of such occurrences, this was found to be due to the Estimote beacon library.

\subsubsection*{Impact Analysis}
While 18,929 BLE-enabled applications may seem like a relatively small number of applications when compared with the initial dataset of 4.6 million+, a single application may correspond to multiple BLE devices, sometimes even millions of devices as is the case with fitness trackers~\cite{IDC:2017:WearablesMarket}. 
For example, even if we consider the slightly restrictive case of unique applications that do not use cryptography with either reads or writes, the cumulative install count is still in excess of 1,005 million. 
This shows that the attack surface is much larger than may be indicated by the number of APKs. 

It is of course a possibility that the data that is read from a BLE peripheral has no impact on user security or privacy (e.g., device battery levels). Understanding the data within APKs would require a more complex static analysis and is left as future work.

\subsection{Case Study: Firmware Update over BLE} \label{subsection:firmware}
When analyzing our results, we found that one of the APKs that was identified as not having application-layer security was designed for use with a fitness tracker from our test device set. 
The tracker is a low-cost model that, based on the install count on Google Play (1,000,000+), appears to be widely used. 
An analysis of the APK suggested that the device used the Nordic BLE chipset, which could be put into the Legacy DFU mode, which does not require the firmware to be signed. 
To exploit this, we developed an APK that, in accordance with the attacks described in Section~\ref{subsection:nordicexperiments}, connects to the device, sends commands to place it in DFU mode, and then writes a new modified firmware to the device without user intervention. 
The updated firmware in this case was a simple, innocuous modification of the original firmware. 
However, given that the device can be configured to receive notifications from other applications, a malicious firmware could be developed in such a way that all notifications (including second-factor authentication SMS messages or end-to-end encrypted messages) are routed to the malicious application that installed the firmware. 
This attack was possible because the BLE peripheral did not verify the firmware nor the source application (via application-layer security). 
We have informed the application developer of the issue (02 Nov 2018), but have received no response as yet.

While our attack was crafted for a specific device, it does demonstrate that attacks against these types of devices are relatively easy.
An attacker could easily embed several firmware images within a single APK to target a range of vulnerable devices. 

\subsection{Limitations} \label{subsection:limitations}
In this section, we outline some limitations, either in our script or due to the inherent nature of our experiments, that may have impacted our results.

\subsubsection*{Unhandled Data Transfer Mechanisms}
As mentioned in Section~\ref{subsection:testing}, BLECryptracer does not analyze data that is written out to file (including shared preferences), or communicated out to a different application, because it is not straightforward (and many times, not possible) to determine how data will be handled once it has been transferred out of the application under analysis.
It is also possible that an application obtains the data to be written to a BLE device from, or forwards the data read from a BLE device to, another entity, such as a remote server.
That is, the Android application could merely act as a ``shuttle" for the data, which means that an analysis of the APK would not show evidence of usage of cryptography libraries. 
However, the transfer of data to/from a remote server does not in itself indicate cryptographically-processed data, as plain-text values can also be transmitted in the same manner. 
We therefore do not analyze instances of data transfers to external entities. 

BLECryptracer also does not handle data transfers between a source and sink when only one of them is processed within an \texttt{Looper} function or when the data is transmitted via messages. However, when we logged instances of where such functions were called during a trace, we found that of the APKs that utilized such data transfer mechanisms, a large percentage were identified as having cryptographic protection via other data flows. In fact, of the 8834 APKs where cryptography was not identified with BLE writes, only 501 APKs interacted with \texttt{Looper} or \texttt{Messenger}, and an even smaller percentage of APKs were affected for BLE reads.

\subsubsection*{Conditional Statements with Backtracing}
When backtracing a register, BLECryptracer stops when it encounters a constant value assignment. However, it is possible that this value assignment occurs within one branch of a conditional jump, which means that another possible value could be contained within another branch further up the instruction list. 
To identify this, the script would have to first trace forward within the instruction list, identify all possible conditional jumps, and then trace back from the register of interest for all branches. 
This would need to be performed for every method that is analyzed and could result in a much longer processing time per APK file, as well as potentially unnecessary overheads.

\section{Related Work} \label{section:relatedwork}
User privacy has received particular attention in the BLE research community because several widely-used BLE devices, such as fitness trackers and continuous glucose monitors, are intended to always be on the user's person, thereby potentially leaking information about the user's whereabouts at all times. 
Some of the research has focused on the threats to privacy based on user location tracking~\cite{Das:2016:Privacy, Fawaz:2016:Ble}, while others explored the possibility of obtaining personal user data from fitness applications or devices~\cite{Cyr:2014:Security,Korolova:2017:CrossApp}.

While our research \textit{is} concerned with data access and user privacy, we focus more on the impact on privacy and security due to how the BLE standard has been implemented in mobile device architectures, as well as how it is applied by application developers, rather than due to individual BLE firmware design.

The work that is most closely related to ours is the research by Naveed, et al., which explored the implications of shared communication channels on Android devices~\cite{Naveed:2014:MisBinding}. 
In their paper, the authors discussed the issue of Classic Bluetooth and NFC channels being shared by multiple applications on the same device. 
They then demonstrated unauthorized data access attacks against (Classic) Bluetooth-enabled medical devices. 
The authors also performed an analysis of 68 Bluetooth-enabled applications that handled private user data, and concluded that the majority of them offered no protection against this attack. 
Finally, they proposed an operating-system level control for mitigating the attack.

Our work specifically targets pairing-protected characteristics on BLE devices, because BLE appears to slowly be replacing Classic Bluetooth in the personal health and home security domains. 
We demonstrate that the BLE data format and access mechanisms enable even easier attacks than in the case of Classic Bluetooth. 
Further, we identify the impact that the new Android permissions model (introduced in Android v6) has had on the user experience and on malicious applications' capabilities. 
We also perform a much larger-scale analysis over 18,900+ Android applications, to determine how prevalent application-layer security is among BLE-enabled applications. 

\section{Conclusions} \label{section:conclusion}
In this paper, we analyze the risks posed to data on Bluetooth Low Energy devices from co-located Android applications. 
We show the conditions under which an unauthorized Android application would be able to access potentially sensitive, pairing-protected data from a BLE peripheral, once a co-located authorized application has paired and bonded with a BLE peripheral, without the user being aware of the access. 
We also show that, in some cases, an unauthorized application may be able to access such protected data with fewer permissions required of it than would be required of an authorized application. 
We then discuss mitigation strategies in terms of the different stakeholders in the BLE ecosystem. 

We present BLECryptracer, an analysis tool for determining the presence of application-layer security with BLE data. We evaluate it against the taint-analysis tool Amandroid, and present the results from executing BLECryptracer against 18,929 BLE-enabled Android APKs. 
Our results suggest that over 45\% of all applications, and about 70\% of ``Medical" applications, do not implement cryptography-based application-layer security for BLE data. 
We also found, among the applications that \textit{do} use cryptographically processed BLE data, several instances of cryptography misuse, such as the use of insecure cipher modes and hard-coded key values. 
We believe that, if this situation does not change, then as more and more sensitive use cases are proposed for BLE, the amount of private or critical data that may be vulnerable to unauthorized access can only increase. 
We hope that our work increases awareness of this issue and prompts changes by application developers and operating system vendors, to lead to improved protection for BLE data.

\section{Availability} \label{subsection:github}
The code for our BLECryptracer tool is available at 
\begin{center}
\url{https://github.com/projectbtle/BLECryptracer} \\
\end{center}
This repository also contains the SHA256 hashes of the APKs in our dataset, and the source/sink files used for the Amandroid analysis. In addition, it contains source code for the benchmarking applications, as well as a comprehensive breakdown of the results per DroidBench category.

{\normalsize \bibliographystyle{acm}
\bibliography{references}}
\appendix
\section*{Appendix A: BLECryptracer Logic}\label{appendix:blecryptracer}
We describe here the basic tracing mechanism employed by BLECryptracer in order to identify the presence of application-layer security for BLE data.

\subsubsection*{Backtracing BLE writes}
BLE writes use one of the \texttt{setValue} methods in Table~\ref{table:setgetmethods} to first set the value that is to be written, before calling the method for performing the actual write. 
BLECryptracer identifies all calls to these methods, and then traces the origins of the data held in the registers that are passed as input to the methods. 

Considering the smali\footnote{Android applications are typically written in Java and converted into Dalvik bytecode. 
The smali format can be considered an ``intermediate" step between the high-level Java source and the bytecode.} code in Figure~\ref{fig:smalisetvalue} as an example, \texttt{setValue} is invoked at Line 13 and is passed two registers as input. 
As \texttt{setValue} is an instance method, the first input, local register \texttt{v3}, holds the \texttt{BluetoothGattCharacteristic} object that the method is invoked on. 
The second input, parameter register \texttt{p2}, holds the data that is to be written to the BLE device, and is the second argument that is passed to the method \texttt{a} (Line 1). 
BLECryptracer identifies \texttt{p2} as the register that holds the data of interest, and traces backward to determine if this data is the result of some cryptographic processing. 
To achieve this, the method(s) within the APK that invoke method \texttt{a} are identified, and the second input to each such method is traced. 
If the BLE data had come from a local register, rather than a parameter register, BLECryptracer would trace back \textit{within} method~\texttt{a}'s instructions, to determine the origin of the data. 
This backtracing is performed until either a crypto-library is referenced, or a \texttt{const-<>} or \texttt{new-array} declaration is encountered (which would indicate that no cryptography is used). 
Note that calls to any method within the crypto-libraries mentioned in Section~\ref{subsection:cryptolibraries} are accepted as evidence of the use of cryptography with BLE data. 
The tool stops processing an APK at the first instance where such a method call is identified.

During execution, the BLECryptracer maintains a list of registers (set within the context of a method) to be traced, for each \texttt{setValue} method call within the application code. 
This initially contains a single entry, which is the input to the \texttt{setValue} method. 
A new register is added to the list if it appears to have tainted the value of any of the registers already in the list. 
This could be due to simple operations such as \texttt{aget}, \texttt{aput} or \texttt{move-<>} (apart from \texttt{move-result} variants), or it could be as a result of a comparison, arithmetic or logic operation (in which case, the register holding the operand on which the operation is performed is added to the trace list). 
Similarly, if a register obtains a value from an instance field (via \texttt{sget} or \texttt{iget}), then all instances where that field is assigned a value are analyzed. 
However, the script does not analyze the order in which the field is assigned values, as this would require activity life-cycle awareness. 

\begin{figure}
	\begin{lstlisting}
.method private a(Landroid/bluetooth/BluetoothGatt;[B...)V
    .locals 10

    .prologue
    const/4 v9, 0x2
    const/4 v8, 0x3
    const/4 v7, 0x1
    ...
    invoke-virtual {v0, v3}, Landroid/bluetooth/BluetoothGattService;-> getCharacteristic(Ljava/util/UUID;) Landroid/bluetooth/BluetoothGattCharacteristic;
   
    move-result-object v3
    ...
    invoke-virtual {v3, p2}, Landroid/bluetooth/BluetoothGattCharacteristic; ->setValue([B)Z
    invoke-virtual {v1, v3}, Landroid/bluetooth/BluetoothGatt; ->writeCharacteristic(Landroid/bluetooth/ BluetoothGattCharacteristic;)Z
	\end{lstlisting}
	\caption{Sample smali code for BLE attribute write.}
	\label{fig:smalisetvalue}
\end{figure}

Where a register is assigned a value that is output from a method invocation via \texttt{move-result}, if the method is not an external method, then the instructions within that method are analysed, beginning with the return value and tracing backwards. 
In some instances, the actual source of a register's value is obfuscated due to the use of intermediate formatting functions. 
In an attempt to overcome this, BLECryptracer traces the inputs to called methods as well. 
Further, if a register is used as input to a method, then all other registers that are inputs to the method are also added to the trace list. 
While this captures some indirect value assignments, it runs the risk of false positives. 
For this reason, we have included the concept of \textit{Confidence Levels} for the code output. 

If, for an APK, the input to the \texttt{setValue} method can be backtraced to cryptography directly, via only register value transfers and as immediate results of method invocations, then a confidence level of ``High" is assigned to the result. 
If a register cannot be traced back directly to a cryptographic output, but if an indirect trace identifies the use of a cryptography library, then a confidence level of ``Medium" is assigned. 
Finally, in the event that no cryptography use is identified at High or Medium confidence levels, the script performs a less stringent search through all the instructions of the methods that it previously analyzed. 
This risks including instances of cryptography use with functions unrelated to BLE and is therefore assigned a ``Low" confidence level.

\begin{figure}
	\centering
	\begin{lstlisting}
.method public onCharacteristicread(Landroid/bluetooth/ BluetoothGatt;Landroid/bluetooth/ BluetoothGattCharacteristic;I)V
    ...
    invoke-virtual {p2}, Landroid/bluetooth/ BluetoothGattCharacteristic;->getValue()[B
    move-result-object v0
    new-instance v2, Ljava/lang/StringBuilder;
    invoke-direct {v2}, Ljava/lang/StringBuilder;-><init>()V
    const-string v3, "read value: "
    invoke-virtual {v2, v3}, Ljava/lang/StringBuilder;->append(Ljava/lang/ String;)Ljava/lang/StringBuilder;
    move-result-object v2
    invoke-static {v0}, Ljava/util/Arrays;->toString([B)Ljava/lang/ String;
    move-result-object v3
    invoke-virtual {v2, v3}, Ljava/lang/StringBuilder;->append(Ljava/lang/ String;)Ljava/lang/StringBuilder;
    move-result-object v2
    ...
	\end{lstlisting}
	\caption{Sample smali code for BLE attribute read.}
	\label{fig:smaligetvalue}
\end{figure}

\subsubsection*{Forward-tracing BLE reads}
With BLE reads, a \texttt{getValue} variant is invoked and the output, i.e., the value that is read, is moved to a register. 
To trace this value, BLECryptracer identifies all calls to \texttt{getValue} variants, then traces the output registers and all registers they taint until either a crypto-library is referenced or the register value changes. 
Such value changes can occur due to \texttt{new-array}, \texttt{new-instance} and \texttt{const} declarations, as well as by being assigned the output of various operations (such as method invocations or arithmetic/logic operations).

With forward-tracing, the register holding the BLE data is considered to taint another if, for example, the source register is used in a method invocation, or comparison/arithmetic/logic operation, whose result is assigned to the destination register. 
The destination register is then added to the trace list. When a register is used as input to a method, then along with the output of that method, the use of the register \textit{within} the method is also analyzed. 

This method of analysis tends to result in a ``tree" of traces. 
As an example, considering the smali code in Figure~\ref{fig:smaligetvalue}, the byte array output from the BLE read is stored in register \texttt{v0} (Line 4). 
This taints register \texttt{v3} via a format conversion function (Lines 10 and 11), which in turn taints \texttt{v2} via a \texttt{java.lang.StringBuilder} function (Lines 12 and 13). 
At this point, all three registers are tainted and will be traced until their values change. 

The forward-tracing mode also assigns one of three confidence levels to its output. 
``High'' is assigned when cryptographically-processed data is identified via the tracing mechanism above; ``Medium'' is when the use of cryptography is identified by tracing classes that implement interfaces. ``Low'' is assigned when a less stringent search through all encountered methods results in identification of a reference to a cryptography library (similar to the backtracing case).

\subsubsection*{Handling obfuscation}
APKs sometimes employ obfuscation techniques to protect against reverse-engineering, and the question then arises as to whether these techniques may impact the results of our analysis. 
We therefore briefly discuss different obfuscation techniques and why they do not impact our tool. 

One of the most common techniques is \textit{identifier renaming}, where identifiers within the code are replaced with short, meaningless names. 
However, because Androguard operates on smali (rather than Java) code, BLECryptracer is able to overcome the challenges posed by this technique. 
\textit{String encryption} is another obfuscation mechanism, but it again does not affect the output of our tool as BLECryptracer does not search for hard-coded strings. 
Further, our tool was verified against a benchmarking application that utilized \textit{reflection}. 
The most complex obfuscation techniques are \textit{packing} and \textit{runtime-based obfuscation}, but these are typically employed by malware. 
Because we are looking for vulnerable (not malicious) applications, we do not consider these techniques. 
Therefore, in general, we believe our analysis to be unaffected by most benign obfuscation mechanisms.

\end{document}